\documentclass{article}
\usepackage{feynmf}
\usepackage{graphicx}
\usepackage{amssymb,amsmath}
\begin{document}
\title{ Radiative Corrections to the $K_{e3}^{\pm}$ Decay Revised }
\author{
 V. Bytev$^1$, E. Kuraev$^1$, \\
 A. Baratt$^2$, J. Thompson$^2$  }
\maketitle
\bigskip

$^1$ Bogoliubov Laboratory of Theoretical Physics, JINR, Dubna, Moscow

region, 141980, Russia \\

$^2$ Department of Physics and Astronomy, University of Pittsburgh,
PA, U.S.A \\


\begin{abstract}
We consider the lowest order radiative corrections
for the decay $K^{\pm}\to \pi^0 e^{\pm} \nu$, usually referred as
$K_{e3}^{\pm}$ decay. This decay is the best way to extract the value of
the $V_{us}$ element of the CKM matrix. The radiative corrections become
crucial if one wants a precise value of $V_{us}$. The existing calculations
were performed in the late 60's \cite{B,G} and are in disagreement. The
calculation in \cite{G} turns out to be ultraviolet cutoff sensitive.
The necessity of
precise knowledge of $V_{us}$ and the  contradiction between the
existing results  constitute the motivation of our paper.

%
%
We remove the ultraviolet cutoff dependence by using A.Sirlin's prescription;
we set it equal to the $W$ mass. We establish the whole character of small
lepton mass dependence based on the renormalization group approach. In
this way we can provide a simple explanation of Kinoshita--Lee--Nauenberg
cancellation
of singularities in the lepton mass terms in the total width and pion spectrum.
We give an explicit evaluation of the structure--dependent photon emission
based on ChPT in the lowest order. We estimate the
accuracy of our results to be at the level of $1\%$.
The corrected total width is $\Gamma=\Gamma_0(1+\delta)$
with $\delta=0.02\pm0.0002$.
Using the  formfactor value
$f_+(0)=0.9842\pm 0.0084$
 calculated in \cite{CKNRT}
leads to $|V_{us}|=0.2172 \pm0.0055$.
\end{abstract}
\newpage

\section{Motivation}

\begin{figure}[htbp]
\begin{fmffile}{virt1}
\begin{fmfgraph*}(100,80) 
 \fmfleft{i1}
 \fmfright{o1,o2,o3}
 \fmfblob{.20w}{v1}
 \fmf{fermion,tension=2.5}{i1,v1}
 \fmf{fermion}{v1,o1}
 \fmf{fermion}{v1,o2}
 \fmf{dashes}{v1,o3}
 \fmflabel{$p$}{i1}
 \fmflabel{$p_e$}{o2}
 \fmflabel{$p_\nu$}{o1}
 \fmflabel{$p'$}{o3}
\end{fmfgraph*}
\hfill
\begin{fmfgraph*}(100,80) 
 \fmfleft{i1}
 \fmfright{o1,o2,o3}
 \fmfblob{.20w}{v3}
 \fmf{plain,tension=3}{i1,v1}
 \fmf{fermion,tension=3}{v1,v2}
 \fmf{plain,tension=3}{v2,v3}
 \fmf{photon,left,tension=0}{v1,v2}
 \fmf{dashes}{v3,o3}
 \fmf{fermion}{v3,o2}
 \fmf{fermion}{v3,o1}
 \fmflabel{$p$}{i1}
 \fmflabel{$p_e$}{o2}
 \fmflabel{$p_\nu$}{o1}
 \fmflabel{$p'$}{o3}
\end{fmfgraph*}
\hfill
\begin{fmfgraph*}(100,80) 
 \fmfleft{i1}
 \fmfright{o1,o2,o3}
 \fmfblob{.20w}{v1}
 \fmf{plain}{v1,v2}
 \fmf{fermion}{v2,v3}
 \fmf{plain}{v3,o2}
 \fmf{photon,left,tension=0}{v2,v3}
 \fmf{dashes,tension=1/4}{v1,o3}
 \fmf{fermion}{i1,v1}
 \fmf{fermion,tension=1/4}{v1,o1}
 \fmflabel{$p$}{i1}
 \fmflabel{$p_e$}{o2}
 \fmflabel{$p_\nu$}{o1}
 \fmflabel{$p'$}{o3}
\end{fmfgraph*}
\\
\parbox[t]{0.3\textwidth}{\centering $a)$}
\hfill
\parbox[t]{0.3\textwidth}{\centering $b)$}
\hfill
\parbox[t]{0.3\textwidth}{\centering $c)$}
\vfill
\end{fmffile}
\end{figure}
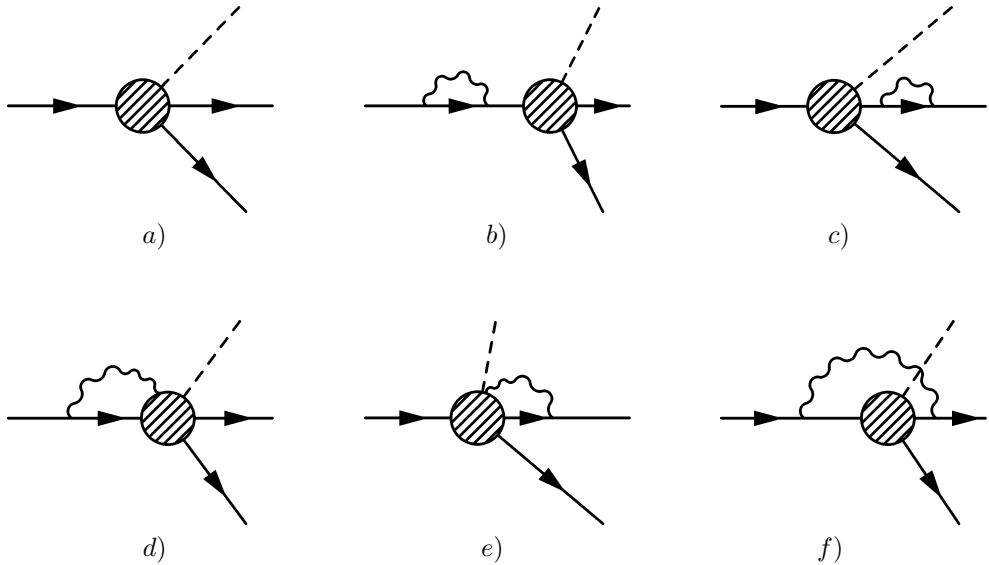
\begin{figure}[htbp]
\begin{fmffile}{virt2}
\begin{fmfgraph*}(100,80) 
 \fmfleft{i1}
 \fmfright{o1,o2,o3}
 \fmfblob{.20w}{v2}
 \fmf{plain,tension=3}{i1,v1}
 \fmf{fermion,tension=2}{v1,v2}
 \fmf{photon,left,tension=0}{v1,v2}
 \fmf{dashes}{v2,o3}
 \fmf{fermion,tension=1/3}{v2,o2}
 \fmf{fermion}{v2,o1}
 \fmflabel{$p$}{i1}
 \fmflabel{$p_e$}{o2}
 \fmflabel{$p_\nu$}{o1}
 \fmflabel{$p'$}{o3}
\end{fmfgraph*}
\hfill
\begin{fmfgraph*}(100,80) 
 \fmfleft{i1}
 \fmfright{o1,o2,o3}
 \fmfblob{.20w}{v1}
 \fmf{fermion}{v1,v3}
 \fmf{plain}{v3,o2}
 \fmf{photon,left,tension=0}{v1,v3}
 \fmf{dashes,tension=1/4}{v1,o3}
 \fmf{fermion}{i1,v1}
 \fmf{fermion,tension=1/4}{v1,o1}
 \fmftop{o3}
 \fmflabel{$p$}{i1}
 \fmflabel{$p_e$}{o2}
 \fmflabel{$p_\nu$}{o1}
 \fmflabel{$p'$}{o3}
\end{fmfgraph*}
\hfill
\begin{fmfgraph*}(100,80) 
 \fmfleft{i1}
 \fmfright{o1,o2,o3}
 \fmfblob{.20w}{v2}
 \fmf{fermion,tension=2}{i1,v1}
 \fmf{plain,tension=2}{v1,v2}
 \fmf{plain,tension=1/2}{v2,v3}
 \fmf{fermion,tension=1/2}{v3,o2}
 \fmf{photon,left,tension=0}{v1,v3}
 \fmf{fermion}{v2,o1}
 \fmf{dashes}{v2,o3}
 \fmflabel{$p$}{i1}
 \fmflabel{$p_e$}{o2}
 \fmflabel{$p_\nu$}{o1}
 \fmflabel{$p'$}{o3}
\end{fmfgraph*}
\\
\parbox[t]{0.3\textwidth}{\centering $d)$}
\hfill
\parbox[t]{0.3\textwidth}{\centering $e)$}
\hfill
\parbox[t]{0.3\textwidth}{\centering $f)$}
\end{fmffile}
\caption{virtual photons}
\label{fig:virt}
\end{figure}
\vfill
%
\begin{figure}[htbp]
\begin{fmffile}{second}
\begin{fmfgraph*}(100,80) 
 \fmfleft{i1}
 \fmfright{o1,o2,o3}
 \fmftop{o4}
 \fmfblob{.20w}{v2}
 \fmf{plain,tension=5}{i1,v1}
 \fmf{fermion,tension=2}{v1,v2}
 \fmf{photon,tension=0}{v1,o4}
 \fmf{dashes,tension=2}{v2,o3}
 \fmf{fermion,tension=1/3}{v2,o2}
 \fmf{fermion,tension=2}{v2,o1}
 \fmflabel{$p$}{i1}
 \fmflabel{$p_e$}{o2}
 \fmflabel{$p_\nu$}{o1}
 \fmflabel{$p'$}{o3}
 \fmflabel{$q$}{o4}
\end{fmfgraph*}
\hfill
\begin{fmfgraph*}(100,80) 
 \fmfleft{i1}
 \fmfright{o1,o2,o4}
 \fmftop{o3}
 \fmfblob{.20w}{v1}
 \fmf{fermion,tension=2}{i1,v1}
 \fmf{plain}{v1,v2}
 \fmf{fermion}{v2,o2}
 \fmf{photon,tension=0}{v2,o4}
 \fmf{dashes,tension=1}{v1,o3}
 \fmf{fermion,tension=1}{v1,o1}
 \fmflabel{$p$}{i1}
 \fmflabel{$p_e$}{o2}
 \fmflabel{$p_\nu$}{o1}
 \fmflabel{$p'$}{o3}
 \fmflabel{$q$}{o4}
\end{fmfgraph*}
\hfill
\begin{fmfgraph*}(100,80) 
 \fmfleft{i1}
 \fmfright{o1,o2,o3}
 \fmftop{o4}
 \fmfblob{.20w}{v1}
 \fmf{fermion,tension=2.5}{i1,v1}
 \fmf{fermion}{v1,o1}
 \fmf{fermion}{v1,o2}
 \fmf{dashes}{v1,o3}
 \fmf{photon,tension=0}{v1,o4}
 \fmflabel{$p$}{i1}
 \fmflabel{$p_e$}{o2}
 \fmflabel{$p_\nu$}{o1}
 \fmflabel{$p'$}{o3}
 \fmflabel{$q$}{o4}
\end{fmfgraph*}
\\
\parbox[t]{0.3\textwidth}{\centering $a)$}
\hfill
\parbox[t]{0.3\textwidth}{\centering $b)$}
\hfill
\parbox[t]{0.3\textwidth}{\centering $c)$}
\end{fmffile}
\caption{real photons}
\label{fig:real}
\end{figure}
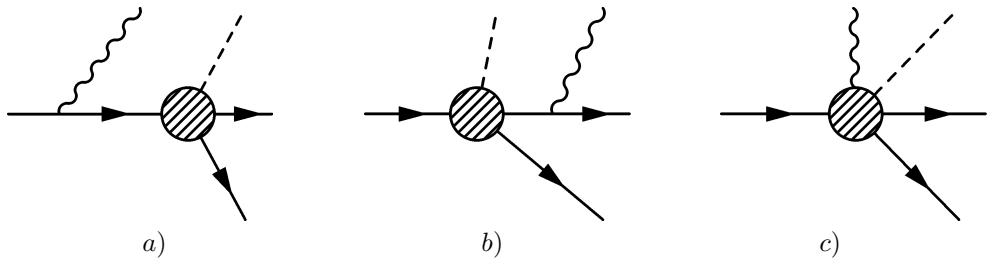

  For corrections due to virtual photons see figure \ref{fig:virt}, for
corrections due to real photons see figure \ref{fig:real}.

  The $K_{e3}$ decay is important since it is the cleanest way to measure the
$V_{us}$ matrix element of the CKM matrix. If one uses the current values for
$V_{ud}$, $V_{us}$, and $V_{ub}$ taken from the PDG then
$|V_{ud}|^2+|V_{us}|^2+|V_{ub}|^2$ misses unity by $2.2$ standard deviations
which contradicts the unitarity of the CKM matrix and might indicate physics
beyond the Standard Model. The uncertainty brought to the above expression by
$V_{us}$ is about the same as uncertainty that comes from $V_{ud}$. Therefore
reducing the error in the $V_{us}$ matrix element would reduce substantially
the error in the whole unitarity equation.
Reliable radiative corrections,    potentially of the
order of a  few percent are necessary  to extract the $V_{us}$
matrix element from the $K_{e3}$ decay width with high precision.

  Calculations of the radiative corrections to the $K_{e3}$ decay were
performed independently by E.S.Ginsberg and T.Becherrawy in the late 60's
\cite{G,B}.
Their results for corrections to the decay rate, Dalitz plot, pion and
positron spectra disagree, in some places quite sharply; for example
Ginsberg's correction to the decay rate is $-0.45\%$ while that of
Becherrawy is $-2\%$ (corresponding to corrections to the
total width $\Gamma$ of 0.45 and 2 respectively).
We have decided to perform a new calculation since results of the  
experiments will become available soon and to explore the causes
of the discrepancies in the previous calculations.
Recently a revision of
E. Ginsberg's  paper, with  numerical estimation of
the radiation corrections
 \cite{CKNRT} was published.

Our paper is organized as follows. The introduction (Section 2) is devoted to the short review
of kinematics of the elastic decay process (without emission of real photon).
In Section 3 we put the results concerning the virtual and soft real photons'
emission contribution to the differential width. In Section 4 we consider the
hard photon emission including both the inner bremsstrahlung (IB) and the
structure-dependent (SD) contributions and   derive an expression
for the differential width
by starting with the  Born width  and adding
the known structure functions in the leading logarithmical
approximation (the so-called Drell-Yan picture of the process).
We give the explicit expressions for the non-leading contributions.
In Section 5, we summarize our results and compare them with those in the  previously published papers.

Appendix A contains the details of calculations of virtual and real soft photons emission.

Appendix B contains the details of description of hard photon emission both by IB and SD mechanisms.
Our approach to study the hard photon emission
differs technically from the ones used in papers \cite{B,G}.

 Appendix C contains the explicit formulae
for description of SD emission including the interference of IB and SD amplitudes.

Appendix D is devoted to analysis of Dalitz-plot distribution and the properties of Drell-Yan conversion mentioned above.

Appendix E contains the list of
the formulae used for the numerical integration.

Appendix F contains the details of kinematics of radiative kaon decay and,besides the analysis of relations of
our and paper \cite{G} technical approaches.

In tables 1,2 and graphics (fig. (3,4,5,6)) the result of numerical estimation of Born values and the correction to
Dalitz-plot distribution and pion and positron spectra are given.

\section{Introduction}

  The lowest order perturbation theory (PT) matrix element of the process
$K^+(p) \to \pi^0(p')+e^+(p_e)+\nu(p_\nu)$ has the form
\begin{equation}
 M=\frac{G_F}{\sqrt{2}}V^*_{us} F_\nu(t)\bar{u}(p_\nu)\gamma_\nu(1+\gamma_5)
 v(p_e)
\end{equation}
where $F_\nu(t)=\frac{1}{\sqrt{2}}(p+p')_{\nu}f_+(t)$.
Dalitz plot density which takes into account the radiative corrections (RC)
of the lowest order PT is
\begin{eqnarray}
 \frac{d^2\Gamma}{dydz}=\mathcal{C} a_0(y,z)(1+\delta(y,z))
 \left(1+\lambda_+\frac{t}{m_\pi^2} \right)^2 = \frac{d^2 \Gamma_0(y,z)}
 {dy dz}(1+\delta(y,z)),  \\ \nonumber
 \Gamma_0=\int\frac{d^2\Gamma_0}{d y d z}d yd z, \quad
  \Gamma=\int\frac{d^2\Gamma}{d y d z}d yd z,
\end{eqnarray}
where the momentum transfer squared between kaon and pion is:
\[
t=(p-p')^2=M_K^2(1+r_\pi-z)=M_K^2R(z) \ .
\]
We accept here the following form for the strong interactions induced
form factor $f_+(t)$:
\begin{equation}
 f_+(t)=f_+(0) \left(1+\lambda_+ \frac{t}{m_\pi^2} \right) \ ,
\end{equation}
according to PDG $\lambda_+=0.0276 \pm 0.0021$. From now on we'll use $M^2$
instead of $M_K^2$.
We define
\begin{equation}
 \mathcal{C}=\frac{M^5 G^2_F|V_{us}|^2}{64\pi^3}|f_+(0)|^2 \ ,
\end{equation}
and
\begin{equation}
 a_0(y,z)=(z+y-1)(1-y)-r_{\pi} + O(r_e) \ .
\end{equation}
 Here we follow the notation of \cite{BCEG}:
\begin{equation}
 r_e \equiv m_e^2/M^2, \hspace{0.2cm} r_\pi \equiv m_\pi^2/M^2 \ ;
\end{equation}
where $m_e$, $m_\pi$, and $M_K$ are the masses of electron, pion, and kaon;
two convenient kinematical variables are
\begin{equation}
 y \equiv 2pp_e/M^2, \hspace{0.2cm} z \equiv 2pp'/M^2 \ .
\end{equation}
In the kaon's rest frame, which we'll imply throughout the paper,
$y$ and $z$ are the energy fractions of the positron and pion:
\begin{equation}
 y=2E_e/M, \hspace{0.2cm} z=2E_\pi/M \ .
\end{equation}
The region of $y,z$-plane where $a_0(y,z)>0$ we will named as a region $D$.
Later, when dealing with real photons we'll also use
\begin{equation}
 x=2\omega/M \ .
\end{equation}
with $\omega$-real photon energy.

 The physical region for $y$ and $z$ (further called $D$ -region) is \cite{BCEG}
\begin{eqnarray}
 2 \sqrt{r_e}  \le y \le 1+r_e-r_\pi \ ,                  \nonumber \\
 F_1(y)-F_2(y) \le z \le F_1(y)+F_2(y) \ ,                \nonumber \\
 F_1(y)=(2-y)(1+r_e+r_\pi-y)/\left[2(1+r_e-y)\right] \ ,  \nonumber \\
 F_2(y)=\sqrt{y^2-4r_e}(1+r_e-r_\pi-y)/\left[2(1+r_e-y)\right] \ ;
\end{eqnarray}
or, equivalently,
\begin{eqnarray}
 2 \sqrt{r_\pi} \le z \le 1+r_\pi-r_e \ ,                 \nonumber \\
 F_3(z)-F_4(z)  \le y \le F_3(z)+F_4(z) \ ,               \nonumber \\
 F_3(z)=(2-z)(1+r_\pi+r_e-z)/[2(1+r_\pi-z)] \ ,           \nonumber \\
 F_4(z)=\sqrt{z^2-4r_\pi}(1+r_\pi-r_e-z)/[2(1+r_\pi-z)] \ .
\end{eqnarray}
For our aims we use the simplified form of physical region
(omitting the terms of the order of $r_e$):
\begin{equation}
 2\sqrt{r_e} \le y \le 1-r_\pi \ ,\hspace{0.2cm}
 c(y) \le z \le 1+r_\pi
\end{equation}
with
\[
 c(y)=1-y+\frac{r_\pi}{1-y}
\]
or,
\begin{equation}
 2\sqrt{r_\pi} \le z \le 1+r_\pi \ , \hspace{0.2cm}
 b_-(z) \le y \le b(z)
\end{equation}
with
\[
 b_-(z)=1-\frac{1}{2}\left(z+\sqrt{z^2-4r_\pi} \right) \ ,
\]
\[
 b(z)=1-\frac{1}{2}\left(z-\sqrt{z^2-4r_\pi} \right) \ .
\]
%

For definiteness we give here the numerical value for Born total width.
It is:
\begin{multline}
\frac{G_F^2M_K^5|V_{us}f_+(0)|^2}{64\pi^3}\int dy\int dz a_0(y,z)(1+\lambda_+\frac{t}{m_\pi^2})^2 \\
=5.36|V_{us}f_+(0)|^2\times 10^{-14} MeV.
\end{multline}
Comparing this value with PDG result:
$(\Gamma_{K_+e3})_{exp}=(2.56\pm 0.03)\times 10^{-15}MeV$
we conclude
\begin{equation}
(V_{us}f_+(0)|)_{\alpha=0}=0.218\pm0.002.
\end{equation}
\section{Virtual and soft real photon emission}

 Taking into account the accuracy level of 0.1\% for determination of
$\rho/\rho_0$ we will drop  terms of order $r_e$.
 We will distinguish 3 kinds of contributions to $\delta$: from emission
of virtual, soft real, and hard real photons in the rest frame of kaon:
$\delta=\delta_V+\delta_S+\delta_H $.
  Standard calculation (see Appendix A for details) allows one to obtain
the following contributions: \\
from the soft real photons
\begin{equation}
\label{ds}
 \delta_S=\frac{\alpha}{\pi}\left\{\left(L_e-2\right)
 \ln \frac{2\Delta\epsilon}{\lambda} + \frac{1}{2} L_e -
 \frac{1}{4} L_e^2 +1 - \frac{\pi^2}{6}\right\} \left(1+O(r_e) \right) \ ,
\end{equation}
from the virtual photons $\delta_V=\delta_C+\delta_{PLM} $ that make up charged fermion's
renormalization, $\delta_C$ (throughout this paper we use Feynman
gauge):
\begin{equation}
\label{dc}
 \delta_C= \frac{\alpha}{2\pi} \left\{ \left[
 - \frac{1}{2}L_\Lambda+ \frac{3}{2} \ln r_e + \ln \frac{M^2}{\lambda^2}-
 \frac{9}{4} \right]+ \left[L_\Lambda+ \ln \frac{M^2}
 {\lambda^2}- \frac{3}{4} \right] \right\} \ ,
\end{equation}
here $L_\Lambda=\ln(\Lambda^2/M^2)$, $\Lambda$ is ultraviolet momentum cutoff,
the first term in the curly braces comes from positron,
the second one from kaon;
and for the diagram in fig1(f) in the point like meson (PLM) approximation, $\delta_{PLM}$:
\begin{multline}
\label{plm}
 \delta_{PLM}= \\
 -\frac{\alpha}{2\pi} \left\{ -L_\Lambda-\frac{1}{2} \ln^2 r_e -
 2 L_e+ \ln \frac{M^2}{\lambda^2} L_e -1+2 \ln^2y+
 2 \ln y+2Li_2(1-y) \right\} \ .
\end{multline}
When these contributions are grouped all together the dependence on $\lambda$
(fictitious "photon mass") disappears. According to Sirlin's prescription \cite{S}
 we set $\Lambda=M_W$. The result can be written in the form:
\begin{multline}
\label{scplm}
 1+\delta_S+\delta_C+\delta_{PLM}= S_W
 [1+\frac{\alpha}{\pi} \biggl[
 \left( L_e-1 \right)\left(\ln \Delta +\frac{3}{4} \right) -\ln \Delta \\
 -\frac{\pi^2}{6}+\frac{3}{4}-Li_2(1-y)-\frac{3}{2}\ln y \biggr]] ,  \qquad
 S_W=1+\frac{3\alpha}{4\pi} L_W.
\end{multline}
In the above equations $L_e=2\ln y +\ln(1/r_e)$, and $L_W=\ln(M_W^2/M^2)$;
$M_W$ is the mass of $W^\pm$, $\Delta=\Delta\epsilon/E_e$,
and $\Delta\epsilon$ is the maximal energy (in the rest frame of kaon)
of a real soft photon. We imply $\Delta\epsilon \ll M/2$. For the details
of eqs (\ref{ds}), (\ref{dc}), (\ref{plm}), and (\ref{scplm}) see Appendix A.


 Contribution from soft photon emission from structure--dependent part
(such as for example, interaction with resonances and intermediate $W^\pm$)
is small, of the order
\[
 \frac{\alpha}{\pi} \frac{\Delta \epsilon}{M} \ll 1 \
\]
and thus is also neglected.

\section{Hard photon emission. Structure function approach}

Next we need to calculate contributions from hard photons. We have to
distinguish between inner bremsstrahlung (IB) and the structure-dependent
(SD) contributions:
 $\delta_H=\delta_{IB}+\delta_{int}+\delta_{SD}$, where $\delta_{int}$
is the interference term between the two.
The terms $\delta_{int}$ and $\delta_{SD}$ are considered in the framework
of the chiral perturbation theory (ChPT) to the orders of $(p^2)$ and $(p^4)$
and find their contribution to be at the level of $0.2\%$ (see appendix C).
\begin{multline}
\label{deltaH}
 \delta_H =\frac{\alpha}{2\pi a_0(y,z)}
 \left\{ \right. \left(L_e-1 \right)\left(\Psi(y,z)-a_0(y,z)
 \left(2\ln \Delta + \frac{3}{2}\right) \right) - \\
  \left. 2a_0(y,z) \ln \frac{b(z)-y}{y\Delta} \right\}+ \delta^{hard} \ ,
\end{multline}
with $\delta^{hard}$ given below.

Extracting the short-distances contributions in form of replacement
$\mathcal{C}\to \mathcal{C}S_W$
it is useful to split $\delta$ (see eq.(2))in the form
\begin{equation}
\label{split}
 \delta(y,z)=\delta_L+\delta_{NL},
\end{equation}
where $\delta_L$ is the leading order contribution, it contains the
'large logarithm' $L_e$ and $\delta_{NL}$ is the non--leading contribution, it contains
the rest of the terms.

$\delta_L$ contains terms from $\delta_C$, $\delta_S$, $\delta_{PLM}$,
and the contribution from  the
collinear configuration of hard IB emission (in
the collinear configuration the angle between the positron and  the
emitted photon  is small). $\delta_L$ turns out to be
\begin{equation}
\label{dlp}
 \delta_L=\frac{\alpha(L_e-1)}{2\pi a_0(y,z)}\Psi(y,z) \ .
\end{equation}
%
First we note that the kinematics of hard photon emission  does not
coincide with  the  elastic process (Region D, the strictly allowed
boundaries of the Dalitz plot). In hard photon emission
an additional region in the  $y,z$ plane, namely $y<b_-(z)$ appears.
The nature of this
phenomenon is the same as the known phenomenon of the  radiative tail in the
process of hadron production at colliding $e^+e^-$ beams.

The quantity $\Psi(y,z)$ has  a  different form for Region  $D$ and
outside it:
\begin{equation}
\Psi(y,z)=\Psi_>(y,z), z>c(y), 2\sqrt{r_e}<y<1-r_\pi \ ;
\end{equation}
and
\begin{equation}
\Psi(y,z)=\Psi_<(y,z), z<c(y), 2\sqrt{r_e}<y<1-\sqrt{r_\pi} \ .
\end{equation}
$\Psi_<(y,z)=0$ when  $ y>1-\sqrt{r_\pi}$ .
Functions $\Psi_<,\Psi_>$ are studied in Appendix D.

$\delta_{NL}$ contains contributions from $\delta_C$, $\delta_{PLM}$,
from SD part of hard photons and from the interference term of SD and IB
parts of hard radiation.
\begin{equation}
\label{nl}
 \delta_{NL}=\frac{\alpha}{\pi}\eta(z,y) \ ,
\end{equation}
where
\begin{equation}
 \frac{\alpha}{\pi} \eta(y,z) = \delta^{hard} +
 \frac{\alpha}{\pi} \left[ \frac{3}{4} -\frac{\pi^2}{6} -Li_2(1-y)-
 \frac{3}{2} \ln y -\ln ((b(z)-y)/y) \right] \ ,
\end{equation}
and for the case when the variables $y,z$ are inside $D$ region:
\begin{equation}
\label{deltahard}
 \delta^{hard}=\frac{\alpha}{2\pi a_0(y,z)}Z_2(y,z) \ ;
\end{equation}
\begin{equation}
\label{Z2}
 Z_2(y,z)=-2 Rphot_{1D}(y,z) + Rphot_{2D}(y,z)+ \int \limits_0^{b(z)-y} dx
 \mathcal{J}(x,y,z)  \ .
\end{equation}
Explicit expressions for $Rphot_{1,2}$ and $\mathcal{J}$ are given in
appendix B.
The Born value and the correction to Dalitz-plot distribution $\Delta(y,z)= \delta(y,z) a_0(y,z)$
is illustrated in tables 1,2.


  We see that the leading  contribution from virtual and soft photon
emission is associated with the so called $\delta$--part  of the evolution
equation kernel:
\begin{equation}
 (\delta_C + \delta_S + \delta_{PLM})^{leading} = \frac{\alpha}{2\pi}
 \left(L_e-1 \right)
 \int \frac{a_0(t,z)}{a_0(y,z)} P_\delta^{(1)}\left(\frac{y}{t}\right)
 \frac{dt}{t}
\end{equation}
where
\begin{equation}
 P_\delta^{(1)}(t)=\delta(1-t)\left(2 \ln \Delta + \frac{3}{2} \right) \ .
\end{equation}
 The contribution of hard photon kinematics in the leading order can be found
with the method of quasireal electrons \cite{BFK} as a convolution of the Born
approximation with the $\theta$--part of evolution equation kernel
$P_\theta(z)$:
\begin{equation}
 \delta_H^{leading} \sim \frac{\alpha}{2\pi} \left(L_e-1 \right)
 \int \frac{dt}{t} \frac{a_0(t,z)}{a_0(y,z)}
 P_\theta^{(1)}\left(\frac{y}{t}\right)
\end{equation}
where
\begin{equation}
 P_\theta^{(1)}(z)=\frac{1+z^2}{1-z}\theta(1-z-\Delta).
\end{equation}
In such a way the whole leading order contribution can be expressed in terms
of convolution of the width in the Born approximation with the whole kernel
of the evolution equation:
\begin{equation}
 P^{(1)}(z)= \lim_{\Delta \to 0} \left( P_\delta^{(1)}(z)+
 P_\theta^{(1)}(z) \right) \ .
\end{equation}


 This approach can be extended to use nonsinglet structure functions $D(t,y)$ \cite{KF}:
\begin{eqnarray}
 d\Gamma^{LO}(y,z)=\int\limits_{max[y,b_-(z)]}^{b(z)} \frac{dt}{t} \ d\Gamma_0
 \left( t,z \right)
 D \left( \frac{y}{t}, \ L_e \right), \ t=x+y \ ,\\ \nonumber
%
\label{D}
 D(z,L)=\delta(1-z)+\frac{\alpha}{2\pi} L P^{(1)}(z)+
 \frac{1}{2!} \left( \frac{\alpha L }{2\pi} \right)^2P^{(2)}(z)+...
 \ , \\ \nonumber
%
%
\label{Pi}
 P^{(i)}(z)=\int\limits_z^1\frac{dx}{x}P^{(1)}(x)P^{(i-1)}(\frac{z}{x}) \ , \
 i=2,3,... \ .
\end{eqnarray}

One can check the validity of the useful relation:
\begin{equation}
\int\limits_0^1 dz P^{(1)}(z)=0 \ .
\end{equation}
  The above makes it easy to see that in the limit
$m_e \to 0$
terms that contain $m_e$ do not contribute
to the total width in correspondence with Kinoshita--Lee--Nauenberg
(KLN) theorem \cite{KLN} as well as with results of E.Ginsberg \cite{G}.
%
%
Keeping in mind the representation
\begin{equation}
\label{psii}
  \Psi(y,z) = \int \limits_{max[y,b_-(z)]}^{b(z)} \frac{dt}{t} \ a_0(t,z) P^{(1)}
  \left( \frac{y}{t} \right) \ ,
\end{equation}
one can get convinced (see Appendix D) that the leading logarithmical  contribution to the
total width as well as one to the pion spectrum is zero due to:
\begin{equation}
\int\limits_{2\sqrt{r_\pi}}^{1+r_\pi} dz\int\limits_0^{b(z)}dy\Psi(y,z)=0.
\end{equation}
Using the general properties of the evolution equations kernels, eq(\ref{D})
one can see that KLN cancellation will take place in all orders of the
perturbation theory.
The spectra in Born approximation are (we omit terms $O(r_e) \sim 10^{-6}$): \\
For pion
\begin{eqnarray}
 &&\frac{1}{\mathcal{C}} \frac{d \Gamma_0}{dz} = \phi_0(z)  \ ,
\\ \nonumber
 &&\phi_0(z)=\left(1+\frac{\lambda_+}{r_\pi}R(z) \right)^2  \int \limits_{b_-(z)}^{b(z)} dy a_0(y,z)=
\left(1+\frac{\lambda_+}{r_\pi}R(z) \right)^2   \frac{1}{6}\left(z^2-4r_\pi \right)^{3/2} \ ,
\end{eqnarray}
and for positron
\begin{multline}
\label{39}
 f(y) = \frac{1}{\mathcal{C}} \frac{d \Gamma_0}{dy} = \\ f_0(y)
 \left[1+\frac{2}{3} \left(\frac{\lambda_+}{r_\pi} \right) \frac{y(1-r_\pi-y)}
 {1-y} + \frac{1}{6} \left(\frac{\lambda_+}{r_\pi} \right)^2
 \frac{y^2(1-r_\pi-y)^2}{(1-y)^2} \right] \ , \\
 f_0(y)=\frac{y^2(1-r_\pi-y)^2}{2(1-y)}.
\end{multline}

The corrected pion spectrum in the inclusive set-up of
experiment when integrating over the whole region for $y$ ($0<y<b(z)$) have a form
$\phi_0(z)+(\alpha/\pi)\phi_1(z)$ with
\begin{multline}
\label{rcpions}
 \phi_1(z)=
 \left(1+\frac{\lambda_+}{r_\pi}R(z)\right)^2
 \biggl[\int\limits_0^{b_-(z)}dy [\Psi_<(y,z)\ln y-
 a_0(y,z)\ln\frac{b(z)-y}{b_-(z)-y}+\\ \frac{1}{2}\tilde{Z}_2(y,z)]+
 \int\limits_{b_-(z)}^{b(z)}dy [\Psi_>(y,z)\ln y+a_0(y,z)Z_1(y,z)+
  \frac{1}{2}Z_2(y,z)] \biggr],
\end{multline}
the quantities $Z_1,\tilde{Z}_2$ explained in Appendix E.
This function do not depend on $\ln(1/r_e)$.
Pion spectrum in the exclusive set-up ($y,z$ in the region $D$) will depend on
$L_e$. It's expression is given in the Appendix E.

Numerical estimation of pion spectrum is illustrated in figure (3,5).

The inclusive positron spectrum with the correction of the lowest order is
$f(y)+(\alpha/\pi) f_1(y)$ with $f(y)$ given above and:
\begin{multline}
\label{rcposis}
 f_1(y)= \frac{1}{2}
 \left( L_e-1 \right)I(y)
 -\int\limits_{c(y)}^{1+r_\pi}a_0(y,z) \left(1+\frac{\lambda_+}{r_\pi} R(z)
 \right)^2 \ln ((b(z)-y)/y) dz+   \\
 \left(\frac{3}{4}-\frac{\pi^2}{6}-\frac{3}{2}\ln y-Li_2(1-y) \right) f(y)
 +
 \frac{1}{2} \int\limits_{c(y)}^{1+r_\pi} Z_2(y,z) \left(1+
 \frac{\lambda_+}{r_\pi} R(z)\right)^2 dz
+ \\
 \theta(1-\sqrt{r_\pi}-y)\int\limits_{2\sqrt{r_\pi}}^{c(y)}
 dz(1+\frac{\lambda_+}{r_\pi}R(z))^2 [(1/2)\tilde{Z}_2-a_0(y,z)
 \ln\frac{b(z)-y}{b_-(z)-y}] ,
\end{multline}
with
\begin{equation}
\label{fullI}
  I(y)=j_0(y)+\left(\frac{\lambda_+}{r_\pi} \right)j_1(y)+
 \left(\frac{\lambda_+}{r_\pi} \right)^2 j_2(y) \ ,
\end{equation}
\begin{multline}
 j_0(y)=\int \limits_y^{1-r_\pi}\frac{d t}{t} \int \limits_{c(t)}^{1+r_\pi}dz
 a_0(t,z)P^{(1)}(\frac{y}{t})= \\
 (2\ln\frac{1-r_\pi-y}{y}+\frac{3}{2})f_0(y)+
 \frac{r_\pi^2(1+y^2)}{2(1-y)}\ln\frac{1-y}{r_\pi}+ \\
 \frac{1}{12}(1-r_\pi-y)[1-5r_\pi-2r_\pi^2+y(4-13r_\pi)-17y^2] \ ;
\end{multline}
explicit expressions for $j_1(y)$ and $j_2(y)$ are given in Appendix D.

Numerical estimation of positron spectrum is illustrated in figure (4,6).

One can check the fulfillment of KLN  
cancellation of singular in the limit $m_e \to 0$ terms
for the total width correction:$\int_0^{1-r_\pi}I(y)dy=0$.
The expression for $j_0(y)$ agree with A(2) from the 1966 year paper of
Eduard Ginsberg \cite{G}.


We put here the general expression for differential width of hard
photon emission which might be useful for construction of Monte Carlo
simulation of real photon emission in $K_{e3}$:
\begin{equation}
 d \Gamma_\gamma^{hard} = d \Gamma_0 \frac{\alpha}{2\pi}\frac{dx}{x}
 \frac{dO_\gamma}{2\pi a_0(y,z)} T \ ,
\end{equation}
with
\begin{equation}
 x=\frac{2\omega}{M} > \frac{2\Delta\epsilon}{M}=
 y\frac{\Delta\epsilon}{E_e}, \
 \frac{\Delta\epsilon}{E_e} \ll 1 \ ;
\end{equation}
and $dO_\gamma$ is an element of photon's solid angle. The quantity
$T$ is explained in Appendix B.

For soft photon emission we have
\begin{equation}
\label{dgammasoft}
 d \Gamma_\gamma^{soft} = d \Gamma_0 \frac{\alpha}{2\pi}\frac{dx}{x}
 \frac{dO_\gamma}{2\pi} \left[ -1-\frac{r_e}{(1-\beta_e C_e)^2}+
 \frac{y}{1-\beta_e C_e} \right], \ x<y\frac{\Delta\epsilon}{E_e} \ .
\end{equation}
Integrating over angles within phase volume of hard photon we obtain the
spectral distribution of radiative kaon decay:
\begin{multline}
\label{47}
 \frac{d \Gamma}{d \Gamma_0 dx}=\frac{\alpha}{2\pi} \frac{1}{a_0(y,z)}
 \biggl[\frac{a_0(x+y,z)}{x(x+y)^2}  
 \left((y^2+(x+y)^2)(L_e-1)+x^2 \right) \\ \nonumber
 -\frac{2}{x}a_0(y,z)-2(\frac{R(z)}{x+y}-y)
  +\mathcal{J}(x,y,z)
  \biggr],   \\ y\Delta<x<b(z)-y  , \qquad
 \Delta=\frac{\Delta \epsilon}{E_e} \ll 1.
\end{multline}

\section{Discussion}

\par Structure--dependent contribution to emission of virtual photons
(see Fig 1 d), e)) can be interpreted as a correction to the strong formfactor
of $K \pi$ transition, $f_+(t)$. We assume that this formfactor can be
extracted from experiment and thus do not consider it. The problem of
calculation of RC to $K_{e3}$ and especially the formfactors in the framework
of CHpT with virtual photons was considered in a recent paper
\cite{CKNRT}.




 As in paper \cite{G} we assume a phenomenological form for the
hadronic contribution to the $K-\pi$ vertex, but here we use
explicitly the dependence of the form factor in the form
\begin{equation}
 f_+(t)=f_+(0) \left(1+\frac{\lambda_+}{r_\pi} R(z)  \right) \ .
\end{equation}
We assume that the effect of higher order ChPT as well as RC to the  
formfactors
can be taken as a multiplicative scaling factor for $f_+(0)$, which we
take from 
of a recent paper \cite{CKNRT}.

We assume such an experiment in which only one positron in the
final state is present, but place  no limits on the number of
photons. The
ratio of the LO contributions in the first order to the Born contribution is
a few percent, and  for the second order it is about
\begin{eqnarray}
 \left( \frac{\alpha L_e}{2 \pi} \right)^2 \le 0.03\% \ .
\end{eqnarray}
Due to non-definite sign structure of the leading logarithm contribution
(see eq(\ref{dlp})) there are regions in the kinematically
allowed area where $|\Psi(y,z)|$ is close to zero. In these regions the
non-leading contributions become dominant.

The contribution of the channels with more than 1 charged lepton in the final
state as well as the vacuum polarization effects in higher orders may be
taken into account by introducing the singlet contribution to the structure
functions. The effect will be at the level of 0.03\% and
we omit them within the precision of our calculation.
%

\par The contribution of the $O(p^4)$ terms \cite{BEG} turns out to be small.
Indeed, one can see that they are of the order
$O(\alpha L^r_9, (\bar{p}/\Lambda_c)^2) \le O(10^{-2}\%)$,
$\Lambda_c=4 \pi F_{\pi} \approx 1.2 GeV$ ($F_\pi=93 MeV$ is the pion life time constant), where $\bar{p}$ is the
characteristic momentum of a final particle in the given reaction,
$\bar{p}^2 \le M^2/16 \sim F^2_\pi$. So the terms of the orders
$O(p^4)$ and $O(p^6)$ can be omitted within the accuracy of
$O \left( \frac{\alpha}{\pi} \times 10^{-2} \right)
\le O \left( 10^{-4} \right)$.

%

Our main results are given in formulae (2,21,22,26-28) for Dalitz plot distribution ;
(38-41) for pion and positron spectra;(47) for hard photon emission;(53) for the
value $|V_{us}|$, in the tables and  figures.
The accuracy of these formulas is determined by the following:
\begin{enumerate}
\item we don't account higher order terms in PT, the ones of the order
  of $(\alpha L_e/\pi)^n, n \ge 2$ which is smaller than $0.03 \%$
\item structure--dependent real hard photon emission contribution to RC
  we estimate to be at the level of $0.0005$.
\item higher order CHPT contributions to the structure dependent part
 are of the order $0.05\%$ \cite{BCEG,BEG}
\end{enumerate}
All the percentages are taken with respect to the Born width. All together
we believe the accuracy of the results to be at the level of
$0.01$.
So the final result of our calculation may be written in the form
\begin{equation}
\label{fr}
 \frac{\Gamma}{\Gamma_0}=\left( 1+\delta(1\pm0.01) \right)
\end{equation}
the terms on the RHS are given in (\ref{split}, \ref{dlp}, \ref{nl}).

Here is the list of improvements comparing with the older calculations
\cite{B,G}:
\begin{enumerate}
\item we eliminate the ultra--violet cutoff dependence by choosing
$\Lambda=M_W$,
\item we describe the dependence on the lepton mass logarithm $L_e$ in all
orders of the perturbation theory and explain why the correction to the total
width does not depend on $m_e$,
\item we treat the strong interaction effects by the means of CHPT in its
lowest order $O(P^2)$ and show that the  next order contribution is
small,
\item we give an explicit formulas for the total differential cross
section
and explicit results for corrections to the Dalitz plot and particle spectra
that might be used in experimental analysis.
\end{enumerate}

In the papers of E.Ginsberg the structure-dependent emission was not 
considered. T.Becherrawy, on the other hand, did include it, and this will 
give rise to differences in the Dalitz plot.  In addition, Ginsberg 
used the proton mass as the momentum cutoff parameter. 

%
%

We do not consider the evolution of coupling constant effects in the regions of
virtual photon momentum modulo square from the quantities of order $M_\rho^2$ up to
$M_Z^2$,which can be taken into account \cite{MS} (and for details see \cite{CKNRT})
 by replacing the quantity $S_W$ by the
$S_{EW}=1+(\alpha/\pi)\ln(M_Z^2/M_\rho^2)=1.0232$.
Taking this replacement into account our result for the correction to the
total
width is
\begin{equation}
\frac{\Gamma}{\Gamma_0}=1+\delta=1.02,
\end{equation}
which results in
\begin{equation}
|V_{us}f_+(0)|=0.214\pm0.002.
\end{equation}
So the correction to the total width is $+2 \%$ while Ginsberg's result
is $-0.45 \%$ and Becherrawy's is $-2 \%$.
Neither Ginsberg nor Becherrawy used  the factor $S_{EW}$,
and this factor (1.023) accounts for most of the difference
between Ginsberg's  and our result.
Electromagnetic corrections become negative and have an order of $10^{-3}$.
The effect of the SD part, which 
E.Ginsberg did not consider, is small, of the order of
0.1\%.

We use  the value of formfactor $f_+(0)=0.9842\pm0.0084$ calculated
in the paper \cite{CKNRT} in the framework  of ChPT, including 
virtual photonic loops
and terms of order $O(p^6)$ of ChPT. To avoid 
double counting we use the mesonic contribution to $f_+^{mes}(0)=1.0002\pm0.0022$
and the $p^6$ terms one $f_+(0)|_{p^6}=-0.016\pm0.0008$). Our final result
is:
\begin{equation}
|V_{us}|=0.21715\pm0.0055.
\end{equation}
In  estimating the uncertainty we take into account the uncertainties
arising from structure-dependent
emission $\pm 0.005$, theoretical errors of order $\pm 0.0003$, experimental error $\pm 0.0022$
and the ChPT error in $p^6$ terms $0.0008$.

In tables 1,2 we give  corrections to the distributions in the Dalitz-plot
$d\Gamma/(dy dz) \sim a_0(y,z)+(\alpha/\pi)\Delta(y,z)$.

In figures (3-6) we illustrate the corrections to the  pion and positron
spectra.
Here we see  qualitative agreement for the positron spectrum and
disagreement with the pion spectrum obtained by E.Ginsberg.

%

\section{Acknowledgements}

We are grateful to S.I.Eidelman for interest to this problem,
 and to E. Swanson for constructive critical discussions.
 One of us (E.K.) is grateful to the Department of Physics and Astronomy of
 the University of Pittsburgh for the warm hospitality during
the last stages of this work, A.Ali for valuable discussion and 
V.N.Samoilov for support.
The work was supported partly by grants RFFI 01-02-17437
and INTAS 00366 and by  USDOE DE FG02 91ER40646.
\newpage
\begin{center}
\begin{tabular}{||c||c|c|c|c|c|c|c|c|c||}
\hline
\hline
  $z/y$& 0.07 &0.15&0.25&0.35&0.45&0.55&0.65&0.75&0.85 \\
\hline
\hline
1.025 &3.83&4.37&3.71&2.01&-0.21&-2.47&-4.31&-5.11&-3.81 \\
\hline
0.975& &3.76&3.49&2.07&0.05&-2.07&-3.83&-4.61&-3.35 \\
\hline
0.925& & &3.26&2.13&0.32&-1.67&-3.35&-4.11&-2.88 \\
\hline
0.875& & &3.04&2.18&0.58&-1.26&-2.86&-3.60&-2.39 \\
\hline
0.825& & & &2.25&0.85&-0.86&-2.37&-3.08&-1.88 \\
\hline
0.775& & & & &1.14&-0.41&-1.83&-2.51&-1.28 \\
\hline
0.725& & & & &1.39&-0.04&-1.39&-2.03&-0.72 \\
\hline
0.675& & & & & &0.38&-0.86&-1.48&  \\
\hline
0.625& & & & & & &-0.35&-0.89& \\
\hline
0.580& & & & & & & &-0.23& \\
\hline
\hline
\end{tabular}
\end{center}
Table 1. Correction to Dalitz plot distribution 
$\Delta(y,z)=a_0(y,z)\delta(y,z)\times 10^{3}$ (see eq (2)).
\vspace{1cm}
\begin{center}
\begin{tabular}{||c||c|c|c|c|c|c|c|c|c||}
\hline
\hline
  $z/y$& 0.07 &0.15&0.25&0.35&0.45&0.55&0.65&0.75&0.85 \\
\hline
\hline
1.025 &0.0084&0.069&0.126&0.163&0.181&0.178&0.156&0.114&0.051 \\
\hline
0.975& &0.026&0.089&0.131&0.153&0.156&0.139&0.101&0.0437 \\
\hline
0.925& & &0.051&0.099&0.126&0.134&0.121&0.088&0.036 \\
\hline
0.875& & &0.014&0.066&0.099&0.111&0.104&0.076&0.029 \\
\hline
0.825& & & &0.0337&0.071&0.089&0.086&0.064&0.021 \\
\hline
0.775& & & & &0.043&0.066&0.069&0.051&0.014 \\
\hline
0.725& & & & &0.016&0.044&0.051&0.039&0.006 \\
\hline
0.675& & & & & &0.021&0.034&0.026&  \\
\hline
0.625& & & & & & &0.016&0.014& \\
\hline
0.580& & & & & & & &0.003& \\
\hline
\hline
\end{tabular}
\end{center}
Table 2.  Dalitz plot distribution in Born approximation $a_0(y,z)$.
\vspace{1cm}

\begin{figure}[htbp]
\label{fig3}
\begin{center}
\fbox{\includegraphics[height=8.0cm]{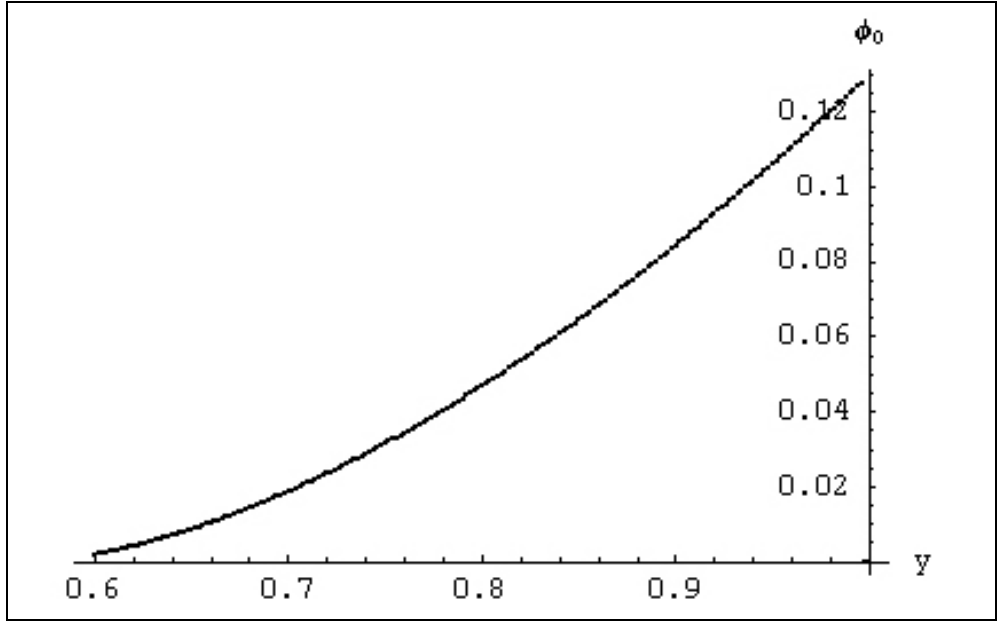}}
\end{center}
\caption{Pion spectrum in Born approximation, $\phi_0(z)$ (see (40)).}
\end{figure}

\begin{figure}[htbp]
\label{fig4}
\begin{center}
\fbox{\includegraphics[height=8.0cm]{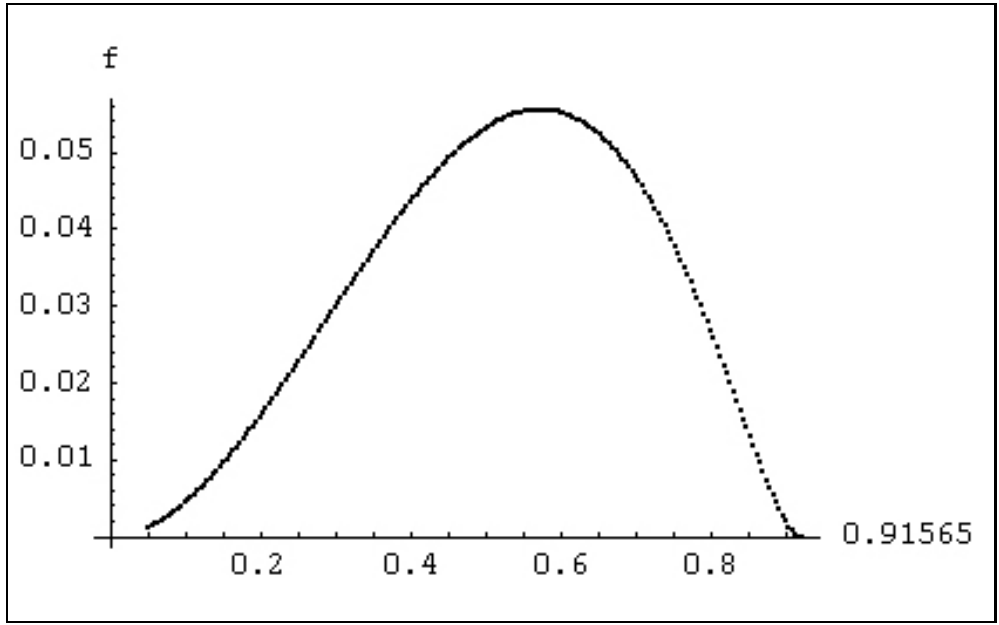}}
\end{center}
\caption{Positron spectrum in Born approximation, $f(y)$ (see (40)).}
\end{figure}

\begin{figure}[htbp]
\label{fig5}
\begin{center}
\fbox{\includegraphics[height=8.0cm]{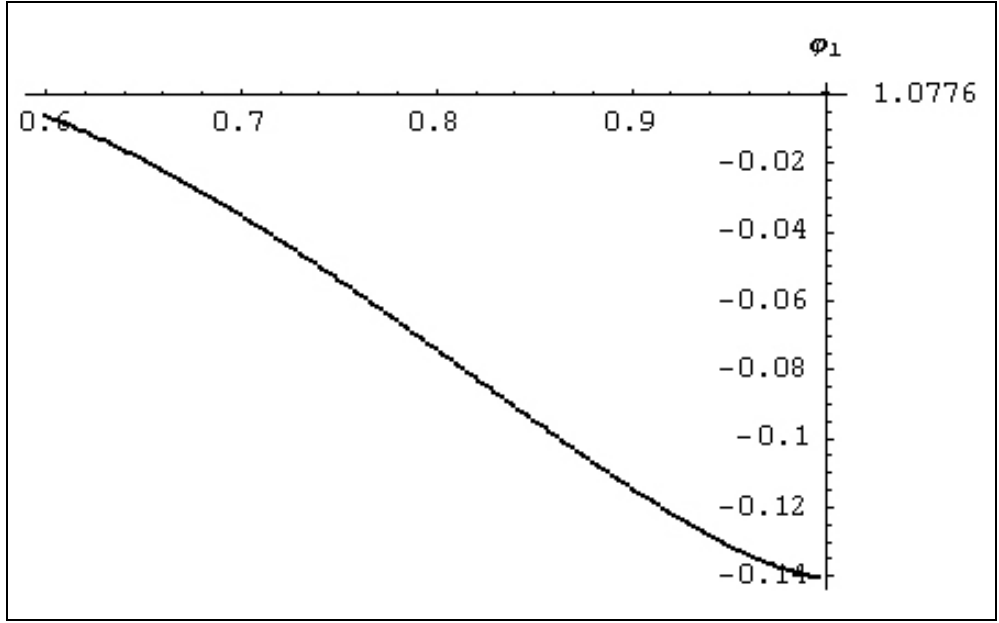}}
\end{center}
\caption{Correction to pion spectrum, $\phi_1(z)$ (see (40)).}
\end{figure}

\begin{figure}[htbp]
\label{fig6}
\begin{center}
\fbox{\includegraphics[height=8.0cm]{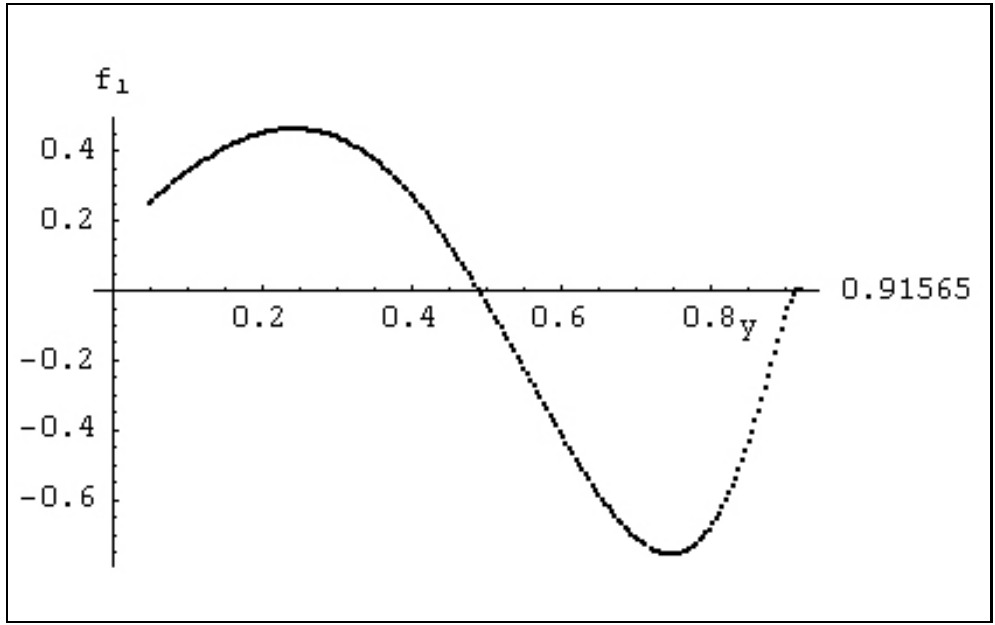}}
\end{center}
\caption{Correction to positron spectrum, $f_1(y)$ (see (41)).}
\end{figure}
%
%
%
%

\section{Appendix A}

\par Here we explain how to calculate $\delta_S$, $\delta_C$,
$\delta_{PLM}$ and how to group them into eq(\ref{scplm}).

Contribution from emission of a soft real photon can be written in
a standard form in terms of the classical currents:
\begin{equation}
 \delta_S=- \frac{4\pi\alpha}{(2\pi)^3} \int \frac{d^3q}{2\omega}
 \left( \frac{p}{p \cdot q}-\frac{p_e}{p_e \cdot q} \right)^2
 \biggm|_{\omega=\sqrt{\bar{q}^2+\lambda^2} < \Delta\epsilon},
\end{equation}
where $\lambda$ is fictitious mass of photon.
We use the following formulas:
\[
 \frac{1}{2\pi} \int \frac{d^3q}{2\omega} \left( \frac{p}{p \cdot q}
 \right)^2=
 \ln \left( \frac{2\Delta\epsilon}{\lambda} \right)-1 \; ;
\]
\[
 \frac{1}{2\pi} \int \frac{d^3q}{2\omega} \left( \frac{p_e}{p_e \cdot q}
 \right)^2=
 \ln \left( \frac{2\Delta\epsilon}{\lambda} \right)-\frac{1}{2}L_e \ ;
\]
\begin{equation}
 \frac{1}{2\pi} \int \frac{d^3q}{2\omega}
 \frac{2(p \cdot p_e)}{(p \cdot q)( p_e \cdot q)} =
 L_e \ln \left( \frac{2\Delta\epsilon}{\lambda} \right)-
 \frac{\pi^2}{6}-\frac{1}{4}L^2_e \ .
\end{equation}
From them we obtain the eq(\ref{ds}).

\par Consider now radiative corrections that arise from emission of
virtual photons (excluding SD virtual photons).

Feynman graphs containing self--energy insertion to positron and kaon
Green functions (Fig.1,b,c) can be taken into account by introducing
the wave function renormalization constants $Z_e$ and $Z_K$:
$M_0 \to M_0(Z_KZ_e)^{1/2}$. We use the expression for $Z_e$ given in the
textbooks \cite{AB}; the expression for $Z_K$ is given in the paper
\cite{Chang}.
The result is eq(\ref{dc}).

Now consider the Feynman graph in which a virtual photon is emitted
by kaon and absorbed by positron or by $W$--boson in the intermediate state
(Fig.1,d,e,f).
This long distance contribution is calculated using a phenomenological model
with point--like mesons as a relevant degrees of freedom. To calculate the
contribution from the region $|k|^2 < \Lambda^2$
($\Lambda$ is the ultra violet cutoff) we use the following expressions
for loop momenta scalar, vector, and tensor integrals:
\begin{equation}
Re \int \frac{d^4 k}{i\pi^2} \frac{1, \ k^\mu, \ k^2}
 {(k^2-\lambda^2)((k-p)^2-M^2)((k-p_e)^2-m_e^2)} = I, \ I^\mu,
 \ J \ .
\end{equation}
A standard calculation yields:
\begin{eqnarray}
 I=\frac{-1}{yM^2} \left\{ \frac{1}{2} \ln \frac{M^2}{\lambda^2}
 L_e+ \ln^2y + Li_2(1-y)- \frac{1}{4} \ln^2r_e \right\} \ ;\\ \nonumber
%
 I^\mu=\frac{-1}{yM^2} \left\{ \frac{-y \ln y}{1-y}p^\mu+
 p_e^\mu \left( \frac{y \ln y}{1-y}+ L_e \right)
 \right\} \ ;\\ \nonumber
%
 J = L_\Lambda + \frac{y \ln y}{1-y} + 1 \ .
\end{eqnarray}
where $L_\Lambda=\ln(\Lambda^2/M^2)$ and we omitted terms of the
order of $O(m_e^2/M^2)$.
As a result we obtain
\begin{multline}
 \int \frac{d^4 k}{i\pi^2} \frac{(1/4)Sp \ p_\nu(p+p')(-\hat{p}_e+\hat{k})
 (2\hat{p}-\hat{k})p_e(p+p')}{(k^2-\lambda^2)((k-p)^2-M^2)
 ((k-p_e)^2-m_e^2)}= 2M^4a_0(y,z)  \times \\
 \left\{ -L_\Lambda-\frac{1}{2} \ln^2 r_e -
 2 L_e+ \ln \frac{M^2}{\lambda^2} L_e -1+2 \ln^2y+
 2 \ln y+2Li_2(1-y) \right\} \ .
\end{multline}

 In a series of papers \cite{S} A.Sirlin has conducted a detailed
analysis of UV behaviour of amplitudes of processes with hadrons
in 1--loop level. He showed that they are UV finite
(if considered on the quark level), but the effective cutoff scale on loop
momenta is of the order $M_W$. For this reason we choose
\[
 L_\Lambda = \ln \frac{M_W^2}{M^2} \ .
\]
The sum $\delta_S+\delta_C+\delta_{PLM}$ yields eq(\ref{scplm})

\section{Appendix B}

The matrix element of the radiative $K_{e3}$ decay
\begin{equation}
K^+(p) \to \pi^0(p')+e^+(p_e)+\nu(p_\nu)+\gamma(q)
\end{equation}
with terms up to $O(p^2)$ in CHPT \cite {GL,BCEG,BEG,H} has the form
\begin{equation}
 M^{hard} = \frac{G}{2} f_+ V_{us}^* \sqrt{4\pi\alpha} \bar{u}(p_\nu)
 Q_\mu^{hard}(1+\gamma_5) v(p_e) \epsilon^\mu(q) \ ,
\end{equation}
where
\begin{eqnarray}
 Q_\mu^{hard} = Q_\mu^e + Q_\mu^\pi + Q_\mu^{SD} = Q_\mu^{IB} + Q_\mu^{SD} \ ;\\ \nonumber
 Q_\mu^{IB}=\left(\hat{p}+\hat{p}'\right) \left[ \frac{(-\hat{p}_e-
 \hat{q}+m_e)\gamma_\mu}
 {2p_eq} +  \frac{p_\mu}{pq} \right] \ ;\\ \nonumber
 Q_\mu^{SD}=\gamma_\nu R_{\mu\nu} \ .
\end{eqnarray}
where tensor $R_{\mu\nu}$ describes (see eq(4.17) in \cite{BCEG}) the
structure--dependent emission (fig 2(c)).
\begin{equation}
  R_{\mu\nu} = g_{\mu\nu} - \frac{q_\nu p_\mu}{pq} \ .
\end{equation}
Singular at $\chi=2p_e q\to 0$ terms which provide contribution containing
large logarithm $L_e$ arises only from
 $Q_\mu^e$. To extract the corresponding terms we
introduce four--vector $v=(x/y)p_e-q$, and $x$ -- is the
energy fraction
of the photon (9). Note that $v \to 0$ when $\chi \to 0$.
Separating leading and non--leading terms and letting $f_+(t)=1$, i.e.
neglecting form factor's momentum dependence, we obtain:
\begin{equation}
 \delta_H=\frac{d\Gamma^{hard}}{d\Gamma_0}=
 \frac{\alpha}{2\pi a_0(y,z)}
 \int  \frac{dx}{x} \int \frac{dO_\gamma}{2\pi}T \ , x>y\Delta \ .
\end{equation}
where
\begin{multline}
\label{T}
 T= \frac{x^2}{8} \sum_{spins} \Bigm|\bar{u}(p_\nu) \left(Q_{IB}^{hard}+
 Q_{SD}^{hard} \right)(1+\gamma_5) v(p_e) \Bigm|^2  = \\
 \frac{y a_0(x+y,z)}{x+y}
 \left[ \frac{y^2+(x+y)^2}{y^2(1-\beta_e C_e)} -
 2\frac{(1-\beta_e)(x+y)}{y(1-\beta_e C_e)^2} \right]-
 \frac{ya_0(x+y,z)}{x+y}+ \mathcal{P} \ .
\end{multline}
The quantity $\mathcal{P}$ contains some non--leading contributions from the
IB part and the ones that arise from the structure--dependent part:
\begin{multline}
 \mathcal{P}= \left( \frac{p_e q}{M^2}
 \left( \frac{p_\nu q}{M^2} + z - \frac{2y}{x+y}\left( 1-x-y\right) \right)
+ \frac{p'v}{M^2} \frac{y(2-x-y)}{x+y} \right) \\
 \left(\frac{xM^2}{4 y p_e q} \left(y^2+(x+y)^2) -1 \right) \right)-
 \frac{M^2x^2}{8p_eq} \left(T_v+\frac{2}{x}T_{1v} \right)-
 \frac{x^2}{8} \left(T_{RR}+2T_R \right) \ ,
\end{multline}
with
\[
  T_v=\frac{1}{4 M^4}Sp \ (\hat{p}+\hat{p}')\hat{p}_\nu(\hat{p}+
  \hat{p}')\hat{v} \ ;
\]
\[
  T_{1v}=\frac{1}{4 M^6}Sp \ (\hat{p}+\hat{p}')
  \hat{p}_\nu(\hat{p}+\hat{p}') \hat{v} \hat{p} \hat{p}_e \ ;
\]
\[
  T_{RR}=R_{\mu\lambda}R_{\mu\sigma}\frac{1}{4M^2}Sp \
  \hat{p}_\nu\gamma_\lambda \hat{p}_e\gamma_\sigma;
\]
\begin{equation}
  T_R=R_{\mu\lambda}\frac{1}{4M^2}Sp \ \hat{p}_\nu(\hat{p}+\hat{p}')
  \left[ \frac{p_\mu}{pq}-
  \frac{(\hat{p}_e+\hat{q})\gamma_\mu}{\chi} \right]
  \hat{p}_e\gamma_\lambda \ .
\end{equation}
To calculate these traces we use the following expressions for the scalar
products of the 4--momenta (in units $M$):
\[
 p^2=1, \hspace{0.25cm} q^2=0, \hspace{0.25cm} p_\nu^2=0, \hspace{0.25cm}
 p^{'2}=r_\pi ,  \hspace{0.25cm} p_e^2 = 0, \hspace{0.25cm}
 p p_e = \frac{y}{2},
\]
\[
 p p'= \frac{z}{2}, \hspace{0.25cm}
 p q= \frac{x}{2}, \hspace{0.25cm}
 p p_\nu= \frac{1}{2}\left( 2-y-z-x \right),
\]
\[
 p' p_\nu = \frac{1}{2} \left(1-x-y-r_\pi+A_e \right), \hspace{0.25cm}
 p' q= \frac{1}{2} \left( x-A_e-A_\nu \right),
\]
\[
 p' p_e =\frac{1}{2} \left(y-R(z)+A_\nu \right), \hspace{0.25cm}
 p_\nu q = \frac{1}{2} A_\nu , \hspace{0.25cm}
 p_e q = \frac{1}{2} A_e ,
\]
\[
 p_e p_\nu = \frac{1}{2} \left(R(z)-A_e-A_\nu \right), \hspace{0.25cm}
 p v =0, \hspace{0.25cm} p_e v =-\frac{1}{2} A_e ,
\]
\[
 q v = \frac{1}{2} \frac{x}{y} A_e , \hspace{0.25cm}
 p'v = \frac{1}{2} \left( \frac{x+y}{y} \tilde{A}_\nu + A_e\right) ,
\]
\[
 p_\nu v = -\frac{1}{2y}\left(x A_e + (x+y) \tilde{A}_\nu \right) \ ,
\]
\[
 \tilde{A}_\nu = A_\nu -\frac{x}{x+y} R(z) \ .
\]
 Three terms in the rhs of (\ref{T}) have a completely different behavior.

The first one corresponds to the kinematic region of collinear emission, when
photon is emitted along positron's momentum. The relevant phase volume has
essentially 3--particles form:
\begin{multline}
 (d \phi_4)^{coll} = \left(\frac{d^3 p_e}{2\epsilon_e}
 \frac{d^3 q}{2 \omega} \frac{d^3 p'}{2 \epsilon'} d^4 p_\nu
 \delta(p_\nu^2)\delta^4(p-p_e-p_\nu-p'-q)  \right)^{coll}= \\
 M^4 \frac{\pi^2}{64} \beta_\pi z dz y dy x dx d O_\gamma d C_{e\pi}
 \delta(1-x-y-z+r_\pi+\frac{x+y}{y} \frac{zy}{2} (1-\beta_\pi C_{e\pi})+
 \frac{2 p_e q}{M^2})= \\
  \frac{y}{x+y} M^4 \frac{\pi^2}{32} dO_\gamma x dx dy dz \ .
\end{multline}
The limits of photon's energy fraction variation are $y\Delta<x<b(z)-y$.
The upper limit is imposed by the Born structure of the width in this
kinematical region.

 The second term corresponds to the contribution from
emission by kaon. The relevant kinematics is isotropic.

The kinematics of radiative kaon decay and the comparison of our and E.Ginsbergs
approaches is given in Appendix F.

 The third term corresponds to the rest
of the contributions which contain neither collinear nor infrared
singularities.
%

 Performing the integration over photon's phase volume provided $y,z$ are
in the $D$ region we obtain:
\begin{multline}
 \int\frac{dx}{x}\int\frac{dO_\gamma}{2\pi}T=\int\limits_{y\Delta}^{b(z)-y}
 \frac{dx}{x} \frac{y^2}{(y+x)^2}a_0(y+x,z)[\frac{y^2+(y+x)^2}{y^2}(L_e-1)+
 \frac{x^2}{y^2}] \ \\
 -2\int\limits_{y\Delta}^{b(z)-y}\frac{dx}{x}[a_0(y,z)+x(\frac{R(z)}{x+y}-y)]+
 \int\limits_0^{b(z)-y} dx\mathcal{J},
\end{multline}
we obtain (\ref{deltahard}) with
\begin{equation}
 Rphot_{1D}=\int\limits_0^{b(z)-y} dx(\frac{R(z)}{x+y}-y)=R(z)\ln\frac{b(z)}{y}-y(b(z)-y);
\end{equation}
\begin{multline}
 Rphot_{_2D}(y,z)=\int\limits_0^{b(z)-y}dx\frac{xa_0(y+x,z)}{(y+x)^2}=-\left(R(z)+y(2-z)\right) \ln \frac{b(z)}{y} + \\ \frac{1}{2}
 \left(b(z)-y \right) \left(2 \frac{R(z)}{b(z)} +4 -2z -b(z) +y \right) \ ,
\end{multline}
and
\begin{equation}
 \mathcal{J}(x,y,z)= \frac{1}{x} \int \frac{dO_\gamma}{2\pi} \mathcal{P} \ .
\end{equation}
One can check that the sum of RC arising from hard, soft and virtual photons
do not depend on the auxiliary parameter $\Delta$.

We note that the leading contribution from hard part of photon spectra can be
reproduced using the method of quasi real electrons \cite{BFK}.
\footnote
{
 the formula (10) in \cite{BFK} should read
\[
  d\Gamma_b = \frac{2\epsilon' d^3 \sigma_{0b}}{d^3 p'}
  \biggm|_{\vec{p}'=\vec{p}_3+\vec{k}} dW_{\vec{p}_3+\vec{k}} \ (k) \
  \frac{d^3p_3}{2\epsilon_3 \ .}
\]
}

Now we concentrate on the contribution of the third term in the RHS of
(\ref{T}).

To perform the integration over the phase volume of final
states it is convenient to use the following parameterization (see Appendix F):
\begin{equation}
 d\phi_4=\frac{d^3p'd^3p_ed^3p_\nu d^3q}
 {2\epsilon' 2E_e 2\epsilon_\nu 2\omega}
 \delta^4 \left( p-p'-p_e-p_\nu-q \right) =
 \beta_\pi \frac{\pi^2}{16}M^4 dy dz x dx
 \frac{d C_e d C_\pi}{\sqrt{D}}  \ ,
\end{equation}
with
\begin{eqnarray}
  D=\beta_\pi^2(1-C^2-C_\pi^2-C_e^2+2C C_\pi C_e) \ , \ \beta_\pi=
  \sqrt{1-\frac{4r_\pi}{z^2}} \ ,\\ \nonumber
 C=\cos(\vec{p}_e,\vec{p'}), \ C_e=\cos(\vec{q},\vec{p}_e), \
 C_\pi=\cos(\vec{q},\vec{p'}) \ .\nonumber
\end{eqnarray}
 The neutrino on--mass shell (NMS) condition provides the relation
\begin{equation}
 1-\beta_\pi C = \frac{2}{yz} \left[ x+y+z-1-r_\pi-\frac{xz}{2}
 \left( 1-\beta_\pi C_\pi \right) -\frac{xy}{2} \left( 1-C_e \right)
 \right] \ .
\end{equation}

 For the aim of further integration of $\mathcal{P}$ over angular variables
we put it in the form:
\begin{eqnarray}
 \mathcal{P}=x P_1 \frac{\tilde{A}_\nu}{A_e} + x P_2 + P_3 A_e + P_4 A_\nu
 + P_5 A_\nu A_e \ , \\ \nonumber
 A_e=\frac{xy}{2}\left(1-C_e \right) \ ,\\ \nonumber
 A_\nu=x-A_e-\frac{xz}{2}\left(1-\beta_\pi C_\pi \right) \ .
\end{eqnarray}
and
\begin{eqnarray}
 P_1 = \frac{y}{2} \left( 1-x-y \right) \ ;\\ \nonumber
 P_2 = \frac{R(z)}{x+y} + \frac{1}{2} \left(z(2x+3y+1)+2x^2+4xy+3y^2
             -2x-3y-2 \right);\\ \nonumber
 P_3 = 1-z-y-\frac{1}{2}x\left(x+y+z \right)  \ ;\\ \nonumber
 P_4 = -1+x+y+\frac{1}{2}xy  \ ;\nonumber
 P_5 = -1 \ .
\end{eqnarray}

Angular integration can be performed explicitly, we have
\begin{equation}
 \int \frac{\beta_\pi dC_\pi}{\pi \sqrt{D}}= \frac{y}{\sqrt{A}} \ ,
 \int \frac{\beta_\pi C_\pi dC_\pi}{\pi \sqrt{D}} =
 \frac{y(x+y-yt)}{z \beta_\pi A^{3/2}}
 \left[ 2R(z)-(x+y)(2-z)+xyt) \right] \ ,
\end{equation}
with
\begin{equation}
 A=(x+y)^2-2xyt, \; t=1-C_e \ .
\end{equation}
Performing the integration over $C_\pi$ we have:
\begin{multline}
 \frac{1}{x}\int \frac{d C_\pi \beta_\pi}{\pi \sqrt{D}}\mathcal{P} =
 \frac{2y}{A^{3/2}} \left( \left(y-x \right)\left(1-\frac{z}{2}-
 \frac{R}{x+y} \right) -\frac{1}{2}y \left(x+y-xt \right) \right) P_1+ \\
 \frac{y}{A^{1/2}}\left( P_2 + \frac{y}{2}t P_3 \right)+
 \left(P_4+\frac{xy}{2}t P_5 \right) \times \\ \left\{ \frac{y}{A^{1/2}}
 \left(1-\frac{z}{2}-\frac{y}{2}t \right)+ \frac{y}{A^{3/2}}
 \left(x+y-yt \right)\left(R-(x+y)\left(1-\frac{z}{2} \right)+\frac{xy}{2}t
 \right) \right\} \ .
\end{multline}
 The following integrals are helpful in integrating the above expression.
We define
\begin{equation}
 I^m_{n}=\int_0^2 \frac{dt t^m}{\sqrt{A^n}}, \ m=0,1,2,3; \ n=1,3.
\end{equation}
Then
\begin{eqnarray}
 I^0_1=\frac{4}{\sigma},\qquad
 I^1_1=\frac{8 (x+y+\sigma)}{3 \sigma^2},\nonumber \\
 I^2_1= \frac{16}{15 \sigma^3} \left(3 \sigma^2+3 (x+y) \rho+5(x+y)^2  \right),\nonumber \\
 I^0_3=\frac{4}{\rho \sigma (x+y)}, \qquad
 I^1_3=\frac{8}{\rho \sigma^2}, \qquad
 I^2_3=\frac{16}{3 \rho \sigma^3} \left(2(x+y)+\sigma \right), \nonumber \\
 I^3_3=\frac{32}{5\rho\sigma^4} \left( \sigma^2+2 (x+y) \rho +4(x+y)^2 \right), \nonumber \\
\end{eqnarray}
where $\rho=|x-y|$ and $\sigma=x+y+\rho$.

The first term in $d\Gamma^{hard}$ together with the leading
contributions from virtual and soft real photons was given in
the form required by RG approach eq(\ref{psii}).

The non--leading contributions, $\delta^{hard}$ from hard photon
emission, include SD emission, IB of point--like mesons as well
as the interference terms. It is free from infrared and mass
singularities and given above (\ref{deltahard}) with
\begin{equation}
 \mathcal{J}(x,y,z)=P_1 R_1 + P_2 y I^0_1 + P_3 \frac{y^2}{2} I^1_1 +
 \frac{y}{2} P_4 R_4 + \frac{xy^2}{4} P_5 R_5 \ ,
\end{equation}
and
\begin{eqnarray}
\nonumber
 R_1 = \frac{y}{x+y} \left(y-x \right)
 \left((2-z)(x+y)-2R(z) \right)I^0_3-y^2((x+y) I^0_3-xI^1_3) \ , \\ \nonumber
%
 R_4 = \left( 2-z\right) I^0_1 - y I^1_1 + (2R(z)-(x+y)(2-z))((x+y)I^0_3 -
 yI^1_3)+ \\ xy((x+y)I^1_3-yI^2_3) \ ,\nonumber \\ \nonumber
%
 R_5 = \left(2-z \right) I^1_1 -yI^2_1 +
 (2R(z)-(x+y)(2-z))((x+y)I^1_3-yI^2_3)+ \\ \nonumber xy((x+y)I^2_3-yI^3_3) \ .
\end{eqnarray}


\section{Appendix C}

 The contribution to $\delta^{hard}$ from SD emission have the form:
\begin{equation}
 \delta_{SD}^{hard}= \frac{\alpha}{2 \pi a_0(y,z)}
 \int \limits_0^{b(z)} dx J^{SD}(x,y,z) \ ,
\end{equation}
where
\begin{equation}
 J^{SD}(x,y,z) = Q_1 R_1 + y Q_2 I^0_1 + \frac{y^2}{2} Q_3 I_1^1 +
 \frac{y}{2} Q_4 R_4 + \frac{x y^2}{4} Q_5 R_5 \ ,
\end{equation}
with $R_i$ given in Appendix B and
\begin{eqnarray}
 Q_1=-\frac{1}{4}y\left(x+y\right) \ ,\nonumber \\
 Q_2=\frac{1}{4} \left[2x(x+2y+z-2)+3y(y+z-2)  \right] \ ,\nonumber \\
 Q_3=-\frac{1}{8} \left[-8+(z+y)(4+3x)-2x+3x^2  \right] \ ,\nonumber \\
 Q_4= \frac{1}{8} \left[ 4y+4x+3xy \right] , \qquad
 Q_5=-\frac{3}{4} \ .
\end{eqnarray}

The contribution to the total width have a form:
\begin{equation}
 \delta^{SD}=\frac{\alpha}{2\pi}\frac{\int\int dy dz
 (1+\frac{\lambda_+}{r_\pi}R(z))^2\int\limits_0^N dx
 J^{SD}(x,y,z)}{\int\int dy dz a_0(y,z)
 (1+\frac{\lambda_+}{r_\pi}R(z))^2}
  \ .
\end{equation}
Numerical estimation gives:
\begin{equation}
 \delta^{SD}=-0.005.
\end{equation}
\section{Appendix D}

The function $\Psi$, defined as
\begin{equation}
\Psi(y,z)=\int\limits_{b_-(z)}^{b(z)}\frac{dt}{t}a_0(t,z)P^{(1)})(\frac{y}{t}),
\end{equation}
contains a restriction on the domain of integration, namely $t$  exceed $y$
or equal to it,which is implied by the kernel $P^{(1)}(y/t)$. Explicit
calculations give:
\begin{multline}
 \Psi_<(y,z)=\int\limits_{b_-(z)}^{b(z)} \frac{dt}{t}a_0(t,z)
 \frac{y^2+t^2}{t(t-y)}=[R(z)-y(2-z)]\ln\frac{b(z)}{b_-(z)}+ \\
 2a_0(y,z)\ln\frac{b(z)-y}{b_-(z)-y}+ \frac{1}{2}(b(z)^2-b_-(z)^2) \ , \\
 2\sqrt{r_e}<y<1-\sqrt{r_\pi},\qquad
 \Psi_<(y,z)=0,  y>1-\sqrt{r_\pi} \ .
\end{multline}
and
\begin{multline}
 \Psi_>(y,z)= \int\limits_y^{b(z)} \frac{dt}{t}a_0(t,z)P^{(1)}(\frac{y}{t})=
 a_0(y,z)[2\ln\frac{b(z)-y}{y}+\frac{3}{2}]- \\
 \frac{1}{2}(b(z)^2-y^2)+
 (b(z)-y)(2-y-z+b_-(z))+[R(z)-y(2-z)]\ln\frac{b(z)}{y}.
\end{multline}
 One can convince the validity of the relations:
\begin{equation}
j_0(y)=\int\limits_{2\sqrt{r_\pi}}^{c(y)}dz\Psi_<(y,z)+\int\limits_{c(y)}^{1+r_\pi}dz\Psi_>(y,z);
\end{equation}
and
\begin{equation}
\int\limits_0^{b_-(z)}dy\Psi_<(y,z)+\int\limits_{b_-(z)}^{b(z)}dy\Psi_>(y,z)=0.
\end{equation}
 The last relation demonstrates the KLN cancellation for the pion spectrum
obtained by integration of the corrections over $y$ in the interval
$0<y<b(z)$.

 The explicit expressions for $j_1(y)$ and $j_2(y)$ are:(for $j_0(y)$ see (\ref{fullI})).
\begin{multline}
 j_1(y)=\frac{y^3(1-r_\pi-y)^3}{3(1-y)^2} \left(2 \ln \frac{1-r_\pi-y}{y}
 + \frac{3}{2} \right) + \\
 \frac{r_\pi^2}{3(1-y)^2} \left[ 3(1-y)(1+y^2)+r_\pi(y^3+3y-2) \right]
 \ln \frac{1-y}{r_\pi} - \\
 \frac{1-r_\pi-y}{36(1-y)^2} \left[ (1-y)^2 \left(43 y^3 -15 y^2 -3y-1 \right.
 +r_\pi(83y^2+26y+11) +3r_\pi^3 \right) + \\ r_\pi^2(31 y^3 -15y^2-39y+47)
  \left.  \right] \ ,
\end{multline}

\begin{multline}
 j_2(y)=\frac{y^4(1-r_\pi-y)^4}{12(1-y)^3} \left(2 \ln \frac{1-r_\pi-y}{y}
 + \frac{3}{2} \right) + \\
 \frac{r_\pi^2}{12(1-y)^3} \ln \frac{1-y}{r_\pi}
 \biggl[ 6(1+y^2)(1-y)^2 -4 r_\pi(1-y)(2y^3-y^2+4y-3)+ \\
 r_\pi^2(y^4+6y^2-8y+3) \biggr] +
 \frac{1-r_\pi-y}{720(1-y)^3} \biggl[-(1-y)^3(247y^4-88y^3-28y^2-8y-3)- \\ 
 r_\pi(1-y)^3(733y^3+341y^2+129y+57)-\\ r_\pi^2(1-y)(707y^4-808y^3+212y^2-
 408y+717)\\ +r_\pi^3(173y^4-72y^3-492y^2+1048y-477)
  -12 r_\pi^4(1-y)^3 \biggr].
\end{multline}

\section{Appendix E}

Collection of the relevant formulae.

The Dalitz-plot distribution in the region $D$:
\begin{multline}
 \frac{1}{\mathcal{C}S_{EW}} \frac{d\Gamma}{dy dz}=\left(1+\lambda_+\frac{t}{m_\pi^2} \right)^2\biggl(a_0(y,z)+
 \frac{\alpha}{\pi}[\frac{1}{2}(L_e-1)
 \Psi_>(y,z)+a_0(y,z)Z_1+\frac{1}{2}Z_2]\biggr),\\
 Z_1=\frac{3}{4}-\frac{\pi^2}{6}-\frac{3}{2}\ln y-
 \ln (({b(z)-y})/y) - Li_2(1-y) \ .
\end{multline}
 The function $Z_2$ is defined in (\ref{Z2}).
 Correction to the total width (we include the contribution of the region outside the region $D$),
 $\Gamma=\Gamma_0(1+\delta)$:
\begin{multline}
 1+\delta=S_{EW}+\frac{\alpha}{\pi} \frac{1}{\int \int dz dy a_0(y,z)
 \left(1+\frac{\lambda_+}{r_\pi}R(z) \right)^2}
 \biggl[ \int\limits_0^{1-r_\pi} I(y)\ln y dy+   \\
  \int\limits_{2\sqrt{r_\pi}}^{1+r_\pi} dz (1+\frac{\lambda_+}{r_\pi}R(z))^2 [\int\limits_0^{b_-(z)}dy[-a_0(y,z)
  \ln\frac{{b(z)-y}}{b_-(z)-y}+(1/2)\tilde{Z}_2(y,z)] + \\
\int\limits_{b_-(z)}^{b(z)}dy[a_0(y,z)Z_1+(1/2)Z_2]\biggr] \ ,
\end{multline}
with
\begin{multline}
 \tilde{Z}_2(y,z)=Rphot_{2A}(y,z)-2Rphot_{1A}(y,z)+\int_{b_-(z)-y}^{b(z)-y} dx \mathcal{J}(x,y,z) ; \\
 Rphot_{1A}(y,z)=R(z)\ln\frac{b(z)}{b_-(z)}-y(b(z)-b_-(z)) \\
 Rphot_{2A}(y,z)=\int_{b_-(z)-y}^{b(z)-y}\frac{ dx x}{(x+y)^2}a_0(x+y,z)= \\
 (b(z)-b_-(z))(1-\frac{z}{2}+2y)-(y(2-z)+R(z))\ln\frac{b(z)}{b_-(z)}.
\end{multline}
The expression in big square brackets in right-hand side of (95) can be put
in the form:
\begin{multline}
\int\limits_{2\sqrt{r_\pi}}^{1+r_\pi}\phi_1(z)dz=\int\limits_{2\sqrt{r_e}}^{1-r_\pi}f_1(y)dy=-0.035
\end{multline}
which results in $\delta=0.02$.
For the aim of comparison with E. Ginsberg result we must put here
\begin{equation}
\lambda_+=0, \qquad I(y)=j_0(y), \qquad M_W=M_p,
\end{equation}
as was mentioned above we have reasonable agreement with E. Ginsberg results.
For the inclusive set-up of experiment (energy fraction of positron is not
measured) we have for pion energy spectrum given above (\ref{rcpions}).
When we restrict ourselves only by the region $D$ the spectrum becomes
dependent on $\ln(1/r_e)$:
\begin{multline}
 \frac{1}{\mathcal{C} S_{EW}}\frac{d\Gamma}{dz}= \left\{ \phi_0(z)+   \right.
 \frac{\alpha}{\pi}[(1/2)P(z)(\ln(1/r_e)-1)+  \\
 \int\limits_{b_-(z)}^{b(z)}dy[\Psi_>(y,z)\ln y+a_0(y,z)Z_1+
 \left. \frac{1}{2}Z_2]] \right\}
 \left(1+\frac{\lambda_+}{r_\pi}R(z) \right)^2 \ ,
\end{multline}
with
\begin{multline}
 P(z)=\frac{1}{6}b_-(z)^2(3b(z)+b_-(z))\ln\frac{b(z)}{b_-(z)}+
 \frac{1}{3}(b(z)-b_-(z))^3\ln\frac{b(z)-b_-(z)}{b(z)}- \\
 \frac{1}{6}b_-(z)(b(z)-b_-(z))(3b_-(z)+b(z)).
\end{multline}
\section{Appendix F}

Our approach to study the radiative kaon decay has an advantage compared
to the one used by E. Ginsberg -- it has a  simple interpretation of
electron mass
singularities based on Drell-Yan picture. The \cite{G} approach to study
 noncollinear
kinematics is more transparent than ours one. We remind the reader of
some some topics of \cite{G}
paper.
One can introduce the missing mass square variable
\begin{equation}
l=(p_\nu+k)^2/M^2=A_\nu=(M-E_\pi-E_e)^2/M^2-(\vec{p}_\pi+\vec{p}_e)^2/M^2.
\end{equation}
the limits of this quantity variation at fixed $y,z$ are pu by the last term:
for collinear or anticollinear kinematics of pion and positron 3-momenta. Being
expressed in terms of $y,z$ they are (we consider the general point of Dalitz-plot and omit positron mass dependence):
\begin{equation}
0<l<b_-(z)(b(z)-y),
\end{equation}
for the $y,z$ in the $D$ region and
\begin{equation}
b(z)(b_-(z)-y)<l<b_-(z)(b(z)-y),
\end{equation}
for the case when they are in the region $A$ outside $D$:
\begin{equation}
0<y<b_-(z),2\sqrt{r_\pi}<z<1+r_\pi.
\end{equation}
For our approach with separating the case of soft and hard photon emission
we must modify the lower bound for $l$ in the region $D$. It can be done
using another representation of $l$:
\begin{equation}
l=x[1-(y/2)(1-C_e)-(z/2)(1-\beta C_\pi)],
\end{equation}
with $C_e,C_\pi$-the cosine of the angles between photon 3-momentum and positron and pion ones, $\beta=\sqrt{1-m^2/E_\pi^2}$-is the pion velocity. Maximum of this quantity is $b(z)$. Taking this into account we obtain for the region oh hard photon
\begin{equation}
x>2\Delta\varepsilon/M=y\Delta,\Delta=\Delta\varepsilon/E_e<<1,
\end{equation}
for the region $D$:
\begin{multline}
y\Delta<x<b(z)-y,\qquad
yb(z)\Delta<l<b_-(z)(b(z)-y);
\end{multline}
and for region $A$:
\begin{multline}
b_-(z)-y<x<{b(z)-y},\qquad
b(z)(b_-(z)-y)<l<b_-(z)(b(z)-y).
\end{multline}
In particular for the collinear case we must choice $C_e=1;C_\pi=-1$,
which corresponds to $x+y<b(z)$. Let infer this condition using the NMS condition :
\begin{multline}
(P_k-p_e-p_\pi-k)^2/M^2=R(z)-x-y+(xy/2)(1-C_e)+\\ (xz/2)(1-\beta C_\pi)+
(yz/2)(1-\beta C_{e\pi})=0.
\end{multline}
In collinear case we have $C_e=1;C_\pi=C_{e\pi}$.From NMS condition we obtain
$1-\beta C_\pi=(2/z(x+y))(x+y-R(z))$. Using this value we obtain $l_{coll}=R(z)x/(x+y)$. Using further the relation $R(z)=b(z)b_-(z)$ we obtain again $x<b(z)-y$ in the case of emission along positron.

Comparing the phase volumes in general case calculated in our approach with
using NMS condition with \cite{G} approach we obtain the relation:
\begin{equation}
 \int x dx\int\frac{dO_\gamma}{4\pi}=\int dl\int d\gamma,
 \int d\gamma=\int\frac{d^3 p_\nu}{E_\nu}\frac{d^3k}{k_0}
 \frac{\delta^4(P-p_\nu-k)}{2\pi}.
\end{equation}
The non--leading contribution arising from hard photon emission considered
above:
\begin{equation}
 I_{IB}=\int\frac{dx}{x}\int\frac{dO_\gamma}{4\pi}\mathcal{P}_{IB},
\end{equation}
with
\begin{multline}
 \mathcal{P}_{IB}=xG_1\frac{\tilde{A}_\nu}{A_e}+xG_2+G_3A_e+G_4A_\nu+
 G_5A_eA_\nu, \\
 G_1=\frac{y}{4}(2-y-x),\\G_2=\frac{R(z)}{2(x+y)}+\frac{x^2}{2}+\frac{1}{2}x(z+2y)+\frac{1}{4}
 (2z+3y(y+z))-1;\\
 G_3=-\frac{1}{8}x^2-\frac{1}{8}x(2+z+y)-\frac{1}{2}(y+z); \qquad
 G_4=\frac{1}{8}x(4+y)+\frac{1}{8}y-1;\\ G_5=-\frac{1}{4},
\end{multline}
(note that $G_i+Q_i=P_i$, see appendix B and C)
can be transformed to the form:
\begin{multline}
 I_{IB}=(1/4)\int dl[4-2y-4z-(1/4)R(z)+(1/4)l+y\ln\frac{(R(z)-l)^2}{l}- \\
2\ln\frac{y^2R(z)^2}{l(l+y(2-z))}+[z+(3/2)y(y+z)-2+(1/4)l(4+y)]I_{10}- \\
(1/2)I_{1-1}-((1/2)l+y+z)I_{2-1}+I_z].
\end{multline}
Here we use the list of integrals obtained in the paper of \cite{G}:
\begin{eqnarray}
 I_{mn}=\int d\gamma\frac{1}{(kP_K/M^2)^m(kp_e/M^2)^n}; \\ \nonumber
 I_{10}=\frac{2}{s}\ln\frac{2-y-z+s}{2-y-z-s}; \qquad I_{20}=4/l; I_{00}=1;\\ \nonumber
 I_{-1,0}=(2-y-z)/2;\qquad I_{11}=\frac{4}{yl}\ln\frac{y^2}{l};\\ \nonumber
 I_{01}=\frac{2}{R(z)-l}\ln\frac{(R(z)-l)^2}{lr_e};\\ \nonumber
 I_{1-1}=\frac{R(z)(2-y-z)-(2+y-z)l}{s^2}+ \\ \nonumber
 \frac{2l(y(2-y-z)-2R(z)+2l)}{s^3}\ln\frac{2-y-z+s}{2-y-z-s};\\ \nonumber
 I_{2-1}=\frac{2(y(2-y-z)+2l-2R(z))}{s^2}+ \\ \nonumber
\frac{R(z)(2-y-z)-(2+y-z)l}{s^3}\ln\frac{2-y-z+s}{2-y-z-s},\\ \nonumber
 s=\sqrt{(2-y-z)^2-4l}.
\end{eqnarray}
Besides we need two additional ones:
\begin{eqnarray}
 I_e=\int d\gamma\frac{1}{(kp_e/M^2)(2(kP_K/M^2)+y}=\frac{2}{yR(z)}\ln
 \frac{y^2R(z)^2}{l(l+y(2-z))r_e}; \\ \nonumber
 I_z=\int d\gamma\frac{1}{(kP_K/M^2)(2(kP_K/M^2)+y}=\frac{4}{ys}\ln
 \frac{2l+ys+y(2-y-z)}{2l+ys-y(2-y-z)}.
\end{eqnarray}
One can see the cancellation of mass singularities
(terms containing $\ln(1/r_e)$) in the expression $I_{IB}$.

Numerical calculations in agreement (within few percent) of this and the
given above expressions.

\end{document}